# The arrival directions of the most energetic cosmic rays

Todor Stanev[1], Peter L. Biermann[2], Jeremy Lloyd–Evans[3],
Jörg P. Rachen[2] and Alan A. Watson[3]

[1] Bartol Research Institute, University of Delaware, Newark, DE 19716, USA
[2] Max Planck Institute for Radioastronomy, D–53010 Bonn, Germany
[3] Department of Physics, University of Leeds, LS2 9JT Leeds, UK

In this Letter we examine the arrival directions of the most energetic cosmic rays ($E > 2 \times 10^{19}$ eV) detected by several air shower experiments. We find that data taken by different air shower arrays show positive correlations, indicating a non–uniform arrival direction distribution. We also find that the events with energy $> 4 \times 10^{19}$ eV exhibit a correlation with the general direction of the supergalactic plane, where a large number of potential sources is located. If confirmed by data from other experiments our results would support models for the extragalactic origin of the highest energy cosmic rays.

The origin of cosmic rays, and especially these of energy above $10^{14}$ eV, is not fully understood. It is difficult to create models for the acceleration of the highest energy ($E > 10^{19}$ eV) particles. Their Larmor radius is so large that they cannot be contained in the Galaxy. On the basis of this argument Cocconi [1] suggested in 1956 that they must come from extragalactic sources. Current models of the extragalactic origin include the acceleration at powerful radio galaxies [2, 3], in clusters of galaxies [4] and the emission of ultra–high energy nucleons from topological defects [5]. The biggest problem with the extragalactic origin of the highest energy events is their large energy loss in interactions with the microwave background which restricts the distance to the potential acceleration sites for any assumption for the type of the primary particle [6, 7, 8, 9]. To avoid this problem others suggest the acceleration in a large galactic halo [10] or even in superbubbles contained in the galactic arms [11]. In this case the highest energy cosmic rays have to be heavy nuclei, with much smaller Lorentz factors and correspondingly smaller gyroradii.

Experimentally it has been very difficult to distinguish between these two extremely different possibilities. Although it has been suggested that the highest energy cosmic rays are less isotropic than the low energy ones [12] the errors on the amplitude and the phase of the anisotropy have not allowed for a more concrete conclusion. One of the reasons is the unknown average value and structure of the extragalactic magnetic field. Charged cosmic rays will scatter on these fields, as well as on the galactic magnetic field, to an unknown angle with the source direction.

Following the observation by two of us [13] that the 8 cosmic rays so far observed above $10^{20}$ eV [14] appear to come from directions close to the supergalactic plane we have undertaken a systematic analysis of 143 events above $2 \times 10^{19}$ eV which have zenith angle, $\theta$, $< 45°$. The choice of the threshold energy of $2 \times 10^{19}$ eV is *a priori* and dictated by the limited statistics. These include the full data set of the Haverah Park experiment [15] and the total available to us of the world data set from the northern hemisphere detectors. We do not use the the data from the University of Sydney experiment, SUGAR [16], because the field of view of that instrument is very different from these of the north hemisphere detectors.

The choice of data with $\theta < 45°$ corresponds to the cut made by the AGASA, Haverah Park and Yakutsk groups when determining the energy spectrum of cosmic rays. The energy of events at large zenith angles is in general less certain because (a) the directional reconstruction is degraded as $\theta$ increases, (b) the uncertainty in the shower attenuation length which is necessarily used to relate showers from different zenith angles to the vertical, and (c) the increasing departure from circular symmetry in the shower plane with angle because of the geomagnetic field [17, 18]

We attempt to measure the average angular distance of the experimentally detected events to the known large scale features of the Galaxy and nearby extragalactic sources – the galactic and supergalactic planes – as a function of the event energy. The supergalactic plane is the symmetry plane of the distribution of cosmologically nearby galaxies.

The Haverah Park set contains 73 such events of energy above $2 \times 10^{19}$ eV. From the AGASA[19] array we have an additional 25 events, and 13 (32) more come from the Volcano Ranch [20] and Yakutsk [18] detectors. These two last event sets are the subsets of the published shower lists [21] with energy above $2 \times 10^{19}$ eV and zenith angle smaller than $45°$. There are slight (order of 25%) systematic differences between the energy estimates in the different experiments and one cannot expect a perfect agreement between different data sets. The full analysis of the arrival distribution has to be performed by the individual experimental groups. We present separately the results for the only complete data set - Haverah Park.

To search for a possible grouping of the experimental data around one of the large scale features we calculated the average and rms angular distance from the galactic (G) and super-galactic (SG) planes : $\langle |b^{G(SG)}| \rangle = \sum_i b_i^{G(SG)}/N$, $b_{RMS}^{G(SG)} = [\sum_i (b_i^{G(SG)})^2/(N-1)]^{1/2}$. To understand the meaning of these experimental quantities we performed a Monte Carlo simulation of uniformly distributed arrival directions by using the declinations of the experimental showers and sampling a random right ascension value. This is necessary as the exposure of each detector, while uniform in right ascension, is a complex function of the galactic latitude and longitude. Only the declinations of the exact experimental set was used. This process results in a loss of sensitivity to any intrinsic anisotropy in declination, but it eliminates the possibility for under(over) estimation of possible energy dependent experimental biases. Ten thousand Monte Carlo sets were simulated for every sample and energy threshold. The probability that the correlation of the experimental data to one of the large scale structures is due to a chance coincidence is estimated by the fraction of the Monte Carlo sets that have $\langle |b| \rangle$ and $b_{RMS}$ value smaller than the experimental one.

The analysis of the correlations between different sets of experimental data was performed by binning of the experimental arrival directions in galactic coordinates. The sky map consists of 125 bins, roughly $18 \times 18°$ each. The bins are not exactly of equal area. The experimental exposure of bin $i$ for each experiment $j$, $\epsilon_i^j$, is simulated by sampling from a uniform distribution, as described above. The probability $P_i$ that $N_i$ or more events are detected in the $i$-th bin by a chance coincidence is calculated for each bin from a cumulative Poissonian distribution. In the case of coincidence between two or more experiments, the chance probabilities were calculated in two different ways. $P_1$ is the probability from a cumulative Poissonian distribution that $\sum N_i$ events are detected for a total exposure of $\sum \epsilon_i$. We also calculated $P_2$ by a direct integration of the joint probability for detection of $N_i$ events by $i$ experiments. $P_1$ and $P_2$ have generally the same statistical significance.

The results from the analysis are shown in Table 1 and in Fig. 1. Table 1 summarizes the analysis of the correlation of the arrival directions of the highest energy cosmic rays with the galactic and supergalactic plane for the Haverah Park and world data set. For the four quantities, $b_{RMS}^G$, $b_{RMS}^{SG}$, $\langle |b^G| \rangle$ and $\langle |b^{SG}| \rangle$ we give the experimental value (*data*), the average from the Monte Carlo sets (*MC*) and the probability $P_u$ that a uniform arrival direction distribution has produced the experimental values. $P_u = P_1$ for all of the available data. Both data sets show the same features – the arrival direction distribution is fully consistent with uniformity at energies above $2 \times 10^{19}$ eV but there is a drastic change when the threshold energy is increased by a factors of 2 and 3. The transition appears to be very sharp. The limited statistics does not allow us to make a more detailed study of the transition region. The magnitude of the effect is 2.5 – 2.8 $\sigma$ in terms of Gaussian probabilities. The large angular spread from the galactic plane, expressed in chance coincidence probabilities very close to 1 (and thus equally statistically unlikely) is a further indication of the non–uniform distribution of the experimental events. Tests performed with the Haverah Park data sets indicate that the dependence of the results on the binning scheme is not strong.

Fig. 1 shows the world statistics above $4 \times 10^{19}$ eV in galactic coordinates. The galactic and supergalactic plane are also shown, together with the exposure of the four experiments for showers with zenith angle less than $45°$. The shaded areas show the regions of the sky where the experimental data above $2 \times 10^{19}$ eV indicate high cosmic ray intensity. Only bins with Poissonian probability ($P_1$) less than 0.1 are plotted. The darkest areas are usually generated by coincidental excesses by two or three experiments.

Table 1: Correlation of the arrival direction of Ultra High Energy cosmic rays with the galactic and the supergalactic plane. The probabilities given do not account for the number of energy bins.

| E>, EeV | # events | $b^{G}_{RMS}$, ° | | | $b^{SG}_{RMS}$, ° | | | $\langle |b^{G}| \rangle$, ° | | | $\langle |b^{SG}| \rangle$, ° | | |
|---|---|---|---|---|---|---|---|---|---|---|---|---|---|
| | | data | MC | $P_u$ | data | MC | $P_u$ | data | MC | $P_u$ | data | MC | $P_u$ |
| 8 events above 100 EeV | | | | | | | | | | | | | |
| 100. | 8 | 47.9 | 40.2 | 0.820 | 26.2 | 38.3 | 0.072 | 35.5 | 31.4 | 0.690 | 18.8 | 30.4 | 0.058 |
| Haverah Park data | | | | | | | | | | | | | |
| 20. | 73 | 39.7 | 36.2 | 0.960 | 32.1 | 32.4 | 0.680 | 33.4 | 29.8 | 0.970 | 26.3 | 26.2 | 0.680 |
| 40. | 27 | 46.5 | 36.7 | 0.996 | 23.3 | 33.5 | 0.006 | 39.5 | 30.2 | 0.995 | 18.6 | 26.9 | 0.013 |
| 60. | 12 | 52.7 | 38.2 | 0.994 | 23.9 | 35.5 | 0.035 | 45.9 | 31.0 | 0.995 | 18.2 | 28.2 | 0.038 |
| All available data | | | | | | | | | | | | | |
| 20. | 143 | 37.9 | 36.7 | 0.890 | 31.9 | 33.5 | 0.340 | 31.2 | 30.2 | 0.880 | 25.5 | 27.3 | 0.310 |
| 40. | 42 | 45.6 | 36.8 | 0.998 | 26.2 | 33.8 | 0.012 | 38.3 | 30.0 | 0.998 | 20.3 | 27.4 | 0.012 |
| 60. | 16 | 48.1 | 38.4 | 0.976 | 26.4 | 36.6 | 0.038 | 41.0 | 31.3 | 0.978 | 19.7 | 29.5 | 0.027 |

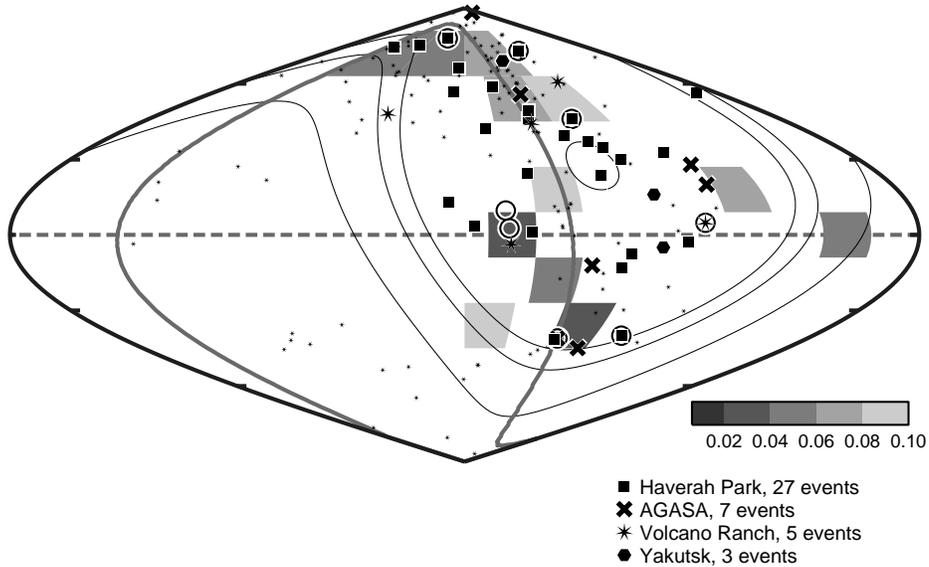

0.02  0.04  0.06  0.08  0.10

■ Haverah Park, 27 events
✕ AGASA, 7 events
✶ Volcano Ranch, 5 events
● Yakutsk, 3 events

Figure 1: Arrival directions of events of energy $> 4 \times 10^{19}$ eV from the four data sets as described in the legend. The events are plotted in galactic coordinates $b_{II}$ and $l_{II}$ centered on the galactic anticenter. $l_{II}$ goes from 0 to 360 degrees from right to left. The 8 events above $10^{20}$ eV are shown with circles. The shading shows the statistical probability ($P_1$) of the detected density of events above $2 \times 10^{19}$ eV by the four detectors. The thin lines show the detector exposures for angle less than 45° – the lower one is for AGASA and Volcano ranch, which are at almost the same latitude (the small cirle is the area around the north celestial pole that they cannot observe), the middle one is for Haverah Park, and the upper one – for Yakutsk. Thick lines show the galactic (dash) and supergalactic planes. Small asterisks show the positions of 118 galaxies within z = 0.005 that are detected in radio from the NASA–IPAC catalog.

The analysis of the events from the Haverah Park experiment and a partial data set available to us from other air shower arrays shows a significant correlation with the direction of the supergalactic plane. The supergalactic plane is a sheet in the sky along which nearby galaxies cluster [22, 23]. Shaver and Pierre [24] extended this to radio galaxies with an approximate distance cutoff of order z = 0.03. Thus radio galaxies are clustered in the sky in this approximate form of a sheet, which runs through the Virgo cluster, through the local group of galaxies, and most other nearby galaxies, such as Cen A, M82, NGC 253, etc. This sheet cuts the galactic plane at nearly 90 degrees close to the anti-center region. Since cosmic rays above $4 \times 10^{19}$ eV are not likely to come from a distance much beyond redshift z = 0.03 due to interaction with the microwave background, they might be expected to show the same clustering [25] if they were associated with these powerful extragalactic sources.

Radio galaxy hot spots have been suggested as sources of the highest energy cosmic rays [2, 3, 26]. These are modelled as very powerful shock waves, with shock velocities which may approach the speed of light. The maximum energies that particles can reach is limited by the space available for Larmor motion in the moving shock frame, $10^{21}$ eV with conventional estimates. Radio galaxies with powerful jets have a magnetic field that is perpendicular to the jet in the outer parts [27]. That means that the maximum energy of a particle fitting into the available space is constant with radial distance in the jet. Assuming that repeated shock structures can compensate adiabatic losses on the way out, particles can also attain very high energies in such jets, not necessarily requiring a hot spot structure. Such a model would increase significantly the number of candidate acceleration sources.

For the upper limit to the intergalactic magnetic field derived by Kronberg [28], of about nanogauss strength with a reversal scale of 1 Mpc, the deflection is of order of only a few degrees at the energies we are interested in here. The apparent anisotropy above $4 \times 10^{19}$ eV indicates that the particle curvature becomes smaller than the angular dimension of the supergalactic plane at that energy. At higher energies the particles would move in a nearly straight path and one could attempt to associate the arrival directions of the highest energy cosmic rays with known astrophysical objects.

We note that the directions in the sky which are most populated in Fig. 1 are close to the directions from which the cosmic ray events of energy above $10^{20}$ eV arrive. These events are shown with large open circles in Fig. 1. It is remarkable that there is an enhancement of lower energy particles in the general direction of each of these events. This is a consistency check, since a source will produce a higher flux at lower energy. Similar clustering around high energy events has been discussed by Chi *et al.* [29] although the clusters so identified lay, very largely, close to the galactic plane.

In the model of radio galaxy origin of the highest energy cosmic rays we suggest the following specific sources:
(a) 3C134, a powerful radio galaxy with hot spots of unknown redshift. It is a viable candidate only if its redshift turns out to be sufficiently small, corresponding to a distance of about 50 Mpc or less.
(b) NGC 315, a well known radio galaxy with powerful asymmetric jets, at redshift 0.0167.
(c) M87 (3C274), also a radio galaxy with a powerful asymmetric jet, in the Virgo cluster, at redshift 0.0043.
(d) Cyg A (3C405), a well known radio galaxy with strong hot spots, although at redshift 0.0565 and thus beyond the local bubble in the galaxy distribution.

All these possible candidates are within about 10° from one or two experimental events of energy above $10^{20}$ eV. There are also other weaker candidates in the neighborhood of some of these sources, such as 3C31 (=NGC 383) at redshift 0.0167 (close to NGC 315). 3C272.1 (=M84) is close to M87 at redshift 0.0031. In terms of sheer radio brightness, Cyg A and M87 are the strongest.

In summary, an analysis of the data available to us indicates that the arrival directions of cosmic rays of energy above $4 \times 10^{19}$ eV are not uniform and that they are concentrated around the direction of the supergalactic plane, which also contains the most of the nearby radiogalaxies. We thus confirm the likely association of some high energy events with the Virgo cluster [31, 32] which is extended along the supergalactic plane. The clustering in the direction of the supergalactic plane is expected for extragalactic models of the origin of the highest energy cosmic rays because it is here that the strong local ($< 100$ Mpc) concentration of powerful radio galaxies, prime candidates [2, 3] for accelerators of the highest energy cosmic rays, occurs. If this correlation is confirmed from analyses of the total world statistics, it would lend considerable support to the hypothesis of powerful radio galaxies as sources for the extreme energy cosmic rays.

It is clearly desirable to make the same test with the full data set from AGASA and Fly's Eye which are not yet available to us. A further test will require construction of the planned 5000 km$^2$ cosmic ray detector, the Auger observatory [33]. Priority should be given to the determination of the redshift of 3C134.

**Acknowledgements.** We wish to thank Prof. M. Nagano for allowing us to use a partial and preliminary data set from the AGASA experiment in this analysis. This paper is a rather direct consequence of the Astrophysics workshop organized by Prof. J.W. Cronin at Fermilab in mid-March 1995. We also acknowledge helpful discussions with Profs. T.K. Gaisser and A.W. Wolfendale. The work of TS is supported in part by U.S. DOE contract. AAW and JLE are supported by UK PPARC.